\documentclass[10pt,conference,twocolumn]{IEEEtran}

\usepackage{cite}
\usepackage[T1]{fontenc}
\usepackage{graphicx}
\usepackage{amssymb}
\usepackage{amsmath}
\usepackage{amsthm}
 
\usepackage{microtype}
\usepackage{balance}
\usepackage{subfigure}
\usepackage{xcolor}
\usepackage{booktabs}
\usepackage{algorithm}
\usepackage{hyperref}

\usepackage{amssymb}
\usepackage{amsfonts}
\usepackage{mathrsfs}
\usepackage{xspace}
\usepackage{bm}
\usepackage{upgreek}

\newcommand{\safemath}[2]{\newcommand{#1}{\ensuremath{#2}\xspace}}

\safemath{\bma}{\mathbf{a}}
\safemath{\bmb}{\mathbf{b}}
\safemath{\bmc}{\mathbf{c}}
\safemath{\bmd}{\mathbf{d}}
\safemath{\bme}{\mathbf{e}}
\safemath{\bmf}{\mathbf{f}}
\safemath{\bmg}{\mathbf{g}}
\safemath{\bmh}{\mathbf{h}}
\safemath{\bmi}{\mathbf{i}}
\safemath{\bmj}{\mathbf{j}}
\safemath{\bmk}{\mathbf{k}}
\safemath{\bml}{\mathbf{l}}
\safemath{\bmm}{\mathbf{m}}
\safemath{\bmn}{\mathbf{n}}
\safemath{\bmo}{\mathbf{o}}
\safemath{\bmp}{\mathbf{p}}
\safemath{\bmq}{\mathbf{q}}
\safemath{\bmr}{\mathbf{r}}
\safemath{\bms}{\mathbf{s}}
\safemath{\bmt}{\mathbf{t}}
\safemath{\bmu}{\mathbf{u}}
\safemath{\bmv}{\mathbf{v}}
\safemath{\bmw}{\mathbf{w}}
\safemath{\bmx}{\mathbf{x}}
\safemath{\bmy}{\mathbf{y}}
\safemath{\bmz}{\mathbf{z}}
\safemath{\bmzero}{\mathbf{0}}
\safemath{\bmone}{\mathbf{1}}

\bmdefine{\biad}{a}
\bmdefine{\bibd}{b}
\bmdefine{\bicd}{c}
\bmdefine{\bidd}{d}
\bmdefine{\bied}{e}
\bmdefine{\bifd}{f}
\bmdefine{\bigd}{g}
\bmdefine{\bihd}{h}
\bmdefine{\biid}{i}
\bmdefine{\bijd}{j}
\bmdefine{\bikd}{k}
\bmdefine{\bild}{l}
\bmdefine{\bimd}{m}
\bmdefine{\bind}{n}
\bmdefine{\biod}{o}
\bmdefine{\bipd}{p}
\bmdefine{\biqd}{q}
\bmdefine{\bird}{r}
\bmdefine{\bisd}{s}
\bmdefine{\bitd}{t}
\bmdefine{\biud}{u}
\bmdefine{\bivd}{v}
\bmdefine{\biwd}{w}
\bmdefine{\bixd}{x}
\bmdefine{\biyd}{y}
\bmdefine{\bizd}{z}

\bmdefine{\bixid}{\xi}
\bmdefine{\bilambdad}{\lambda}
\bmdefine{\bimud}{\mu}
\bmdefine{\bithetad}{\theta}
\bmdefine{\biphid}{\phi}
\bmdefine{\bideltad}{\delta}

\safemath{\bmia}{\biad}
\safemath{\bmib}{\bibd}
\safemath{\bmic}{\bicd}
\safemath{\bmid}{\bidd}
\safemath{\bmie}{\bied}
\safemath{\bmif}{\bifd}
\safemath{\bmig}{\bigd}
\safemath{\bmih}{\bihd}
\safemath{\bmii}{\biid}
\safemath{\bmij}{\bijd}
\safemath{\bmik}{\bikd}
\safemath{\bmil}{\bild}
\safemath{\bmim}{\bimd}
\safemath{\bmin}{\bind}
\safemath{\bmio}{\biod}
\safemath{\bmip}{\bipd}
\safemath{\bmiq}{\biqd}
\safemath{\bmir}{\bird}
\safemath{\bmis}{\bisd}
\safemath{\bmit}{\bitd}
\safemath{\bmiu}{\biud}
\safemath{\bmiv}{\bivd}
\safemath{\bmiw}{\biwd}
\safemath{\bmix}{\bixd}
\safemath{\bmiy}{\biyd}
\safemath{\bmiz}{\bizd}

\safemath{\bmxi}{\bixid}
\safemath{\bmlambda}{\bilambdad}
\safemath{\bmmu}{\bimud}
\safemath{\bmtheta}{\bithetad}
\safemath{\bmphi}{\biphid}
\safemath{\bmdelta}{\bideltad}

\safemath{\bA}{\mathbf{A}}
\safemath{\bB}{\mathbf{B}}
\safemath{\bC}{\mathbf{C}}
\safemath{\bD}{\mathbf{D}}
\safemath{\bE}{\mathbf{E}}
\safemath{\bF}{\mathbf{F}}
\safemath{\bG}{\mathbf{G}}
\safemath{\bH}{\mathbf{H}}
\safemath{\bI}{\mathbf{I}}
\safemath{\bJ}{\mathbf{J}}
\safemath{\bK}{\mathbf{K}}
\safemath{\bL}{\mathbf{L}}
\safemath{\bM}{\mathbf{M}}
\safemath{\bN}{\mathbf{N}}
\safemath{\bO}{\mathbf{O}}
\safemath{\bP}{\mathbf{P}}
\safemath{\bQ}{\mathbf{Q}}
\safemath{\bR}{\mathbf{R}}
\safemath{\bS}{\mathbf{S}}
\safemath{\bT}{\mathbf{T}}
\safemath{\bU}{\mathbf{U}}
\safemath{\bV}{\mathbf{V}}
\safemath{\bW}{\mathbf{W}}
\safemath{\bX}{\mathbf{X}}
\safemath{\bY}{\mathbf{Y}}
\safemath{\bZ}{\mathbf{Z}}

\safemath{\bZero}{\mathbf{0}}
\safemath{\bOne}{\mathbf{1}}
\safemath{\bDelta}{\mathbf{\Delta}}
\safemath{\bLambda}{\mathbf{\UpLambda}}
\safemath{\bPhi}{\mathbf{\Upphi}}
\safemath{\bSigma}{\mathbf{\Upsigma}}
\safemath{\bOmega}{\mathbf{\Upomega}}
\safemath{\bTheta}{\mathbf{\Uptheta}}

\bmdefine{\biAd}{A}
\bmdefine{\biBd}{B}
\bmdefine{\biCd}{C}
\bmdefine{\biDd}{D}
\bmdefine{\biEd}{E}
\bmdefine{\biFd}{F}
\bmdefine{\biGd}{G}
\bmdefine{\biHd}{H}
\bmdefine{\biId}{I}
\bmdefine{\biJd}{J}
\bmdefine{\biKd}{K}
\bmdefine{\biLd}{L}
\bmdefine{\biMd}{M}
\bmdefine{\biOd}{N}
\bmdefine{\biPd}{O}
\bmdefine{\biQd}{P}
\bmdefine{\biRd}{R}
\bmdefine{\biSd}{S}
\bmdefine{\biTd}{T}
\bmdefine{\biUd}{U}
\bmdefine{\biVd}{V}
\bmdefine{\biWd}{W}
\bmdefine{\biXd}{X}
\bmdefine{\biYd}{Y}
\bmdefine{\biZd}{Z}

\bmdefine{\biDelta}{\Delta}
\bmdefine{\biLambda}{\Lambda}
\bmdefine{\biPhi}{\Phi}
\bmdefine{\biSigma}{\Sigma}
\bmdefine{\biOmega}{\Omega}
\bmdefine{\biTheta}{\Theta}

\safemath{\bimA}{\biAd}
\safemath{\bimB}{\biBd}
\safemath{\bimC}{\biCd}
\safemath{\bimD}{\biDd}
\safemath{\bimE}{\biEd}
\safemath{\bimF}{\biFd}
\safemath{\bimG}{\biGd}
\safemath{\bimH}{\biHd}
\safemath{\bimI}{\biId}
\safemath{\bimJ}{\biJd}
\safemath{\bimK}{\biKd}
\safemath{\bimL}{\biLd}
\safemath{\bimM}{\biMd}
\safemath{\bimN}{\biNd}
\safemath{\bimO}{\biOd}
\safemath{\bimP}{\biPd}
\safemath{\bimQ}{\biQd}
\safemath{\bimR}{\biRd}
\safemath{\bimS}{\biSd}
\safemath{\bimT}{\biTd}
\safemath{\bimU}{\biUd}
\safemath{\bimV}{\biVd}
\safemath{\bimW}{\biWd}
\safemath{\bimX}{\biXd}
\safemath{\bimY}{\biYd}
\safemath{\bimZ}{\biZd}

\safemath{\bimDelta}{\biDelta}
\safemath{\bimLambda}{\biLambda}
\safemath{\bimPhi}{\biPhi}
\safemath{\bimSigma}{\biSigma}
\safemath{\bimOmega}{\biOmega}
\safemath{\bimTheta}{\biTheta}

\safemath{\setA}{\mathcal{A}}
\safemath{\setB}{\mathcal{B}}
\safemath{\setC}{\mathcal{C}}
\safemath{\setD}{\mathcal{D}}
\safemath{\setE}{\mathcal{E}}
\safemath{\setF}{\mathcal{F}}
\safemath{\setG}{\mathcal{G}}
\safemath{\setH}{\mathcal{H}}
\safemath{\setI}{\mathcal{I}}
\safemath{\setJ}{\mathcal{J}}
\safemath{\setK}{\mathcal{K}}
\safemath{\setL}{\mathcal{L}}
\safemath{\setM}{\mathcal{M}}
\safemath{\setN}{\mathcal{N}}
\safemath{\setO}{\mathcal{O}}
\safemath{\setP}{\mathcal{P}}
\safemath{\setQ}{\mathcal{Q}}
\safemath{\setR}{\mathcal{R}}
\safemath{\setS}{\mathcal{S}}
\safemath{\setT}{\mathcal{T}}
\safemath{\setU}{\mathcal{U}}
\safemath{\setV}{\mathcal{V}}
\safemath{\setW}{\mathcal{W}}
\safemath{\setX}{\mathcal{X}}
\safemath{\setY}{\mathcal{Y}}
\safemath{\setZ}{\mathcal{Z}}
\safemath{\emptySet}{\varnothing}

\safemath{\colA}{\mathscr{A}}
\safemath{\colB}{\mathscr{B}}
\safemath{\colC}{\mathscr{C}}
\safemath{\colD}{\mathscr{D}}
\safemath{\colE}{\mathscr{E}}
\safemath{\colF}{\mathscr{F}}
\safemath{\colG}{\mathscr{G}}
\safemath{\colH}{\mathscr{H}}
\safemath{\colI}{\mathscr{I}}
\safemath{\colJ}{\mathscr{J}}
\safemath{\colK}{\mathscr{K}}
\safemath{\colL}{\mathscr{L}}
\safemath{\colM}{\mathscr{M}}
\safemath{\colN}{\mathscr{N}}
\safemath{\colO}{\mathscr{O}}
\safemath{\colP}{\mathscr{P}}
\safemath{\colQ}{\mathscr{Q}}
\safemath{\colR}{\mathscr{R}}
\safemath{\colS}{\mathscr{S}}
\safemath{\colT}{\mathscr{T}}
\safemath{\colU}{\mathscr{U}}
\safemath{\colV}{\mathscr{V}}
\safemath{\colW}{\mathscr{W}}
\safemath{\colX}{\mathscr{X}}
\safemath{\colY}{\mathscr{Y}}
\safemath{\colZ}{\mathscr{Z}}

\safemath{\opA}{\mathbb{A}}
\safemath{\opB}{\mathbb{B}}
\safemath{\opC}{\mathbb{C}}
\safemath{\opD}{\mathbb{D}}
\safemath{\opE}{\mathbb{E}}
\safemath{\opF}{\mathbb{F}}
\safemath{\opG}{\mathbb{G}}
\safemath{\opH}{\mathbb{H}}
\safemath{\opI}{\mathbb{I}}
\safemath{\opJ}{\mathbb{J}}
\safemath{\opK}{\mathbb{K}}
\safemath{\opL}{\mathbb{L}}
\safemath{\opM}{\mathbb{M}}
\safemath{\opN}{\mathbb{N}}
\safemath{\opO}{\mathbb{O}}
\safemath{\opP}{\mathbb{P}}
\safemath{\opQ}{\mathbb{Q}}
\safemath{\opR}{\mathbb{R}}
\safemath{\opS}{\mathbb{S}}
\safemath{\opT}{\mathbb{T}}
\safemath{\opU}{\mathbb{U}}
\safemath{\opV}{\mathbb{V}}
\safemath{\opW}{\mathbb{W}}
\safemath{\opX}{\mathbb{X}}
\safemath{\opY}{\mathbb{Y}}
\safemath{\opZ}{\mathbb{Z}}
\safemath{\opZero}{\mathbb{O}}
\safemath{\identityop}{\opI}

\safemath{\veca}{\bma}
\safemath{\vecb}{\bmb}
\safemath{\vecc}{\bmc}
\safemath{\vecd}{\bmd}
\safemath{\vece}{\bme}
\safemath{\vecf}{\bmf}
\safemath{\vecg}{\bmg}
\safemath{\vech}{\bmh}
\safemath{\veci}{\bmi}
\safemath{\vecj}{\bmj}
\safemath{\veck}{\bmk}
\safemath{\vecl}{\bml}
\safemath{\vecm}{\bmm}
\safemath{\vecn}{\bmn}
\safemath{\veco}{\bmo}
\safemath{\vecp}{\bmp}
\safemath{\vecq}{\bmq}
\safemath{\vecr}{\bmr}
\safemath{\vecs}{\bms}
\safemath{\vect}{\bmt}
\safemath{\vecu}{\bmu}
\safemath{\vecv}{\bmv}
\safemath{\vecw}{\bmw}
\safemath{\vecx}{\bmx}
\safemath{\vecy}{\bmy}
\safemath{\vecz}{\bmz}

\safemath{\veczero}{\bmzero}
\safemath{\vecone}{\bmone}
\safemath{\vecxi}{\bmxi}
\safemath{\veclambda}{\bmlambda}
\safemath{\vecmu}{\bmmu}
\safemath{\vectheta}{\bmtheta}
\safemath{\vecphi}{\bmphi}
\safemath{\vecdelta}{\bmdelta}

\safemath{\matA}{\bA}
\safemath{\matB}{\bB}
\safemath{\matC}{\bC}
\safemath{\matD}{\bD}
\safemath{\matE}{\bE}
\safemath{\matF}{\bF}
\safemath{\matG}{\bG}
\safemath{\matH}{\bH}
\safemath{\matI}{\bI}
\safemath{\matJ}{\bJ}
\safemath{\matK}{\bK}
\safemath{\matL}{\bL}
\safemath{\matM}{\bM}
\safemath{\matN}{\bN}
\safemath{\matO}{\bO}
\safemath{\matP}{\bP}
\safemath{\matQ}{\bQ}
\safemath{\matR}{\bR}
\safemath{\matS}{\bS}
\safemath{\matT}{\bT}
\safemath{\matU}{\bU}
\safemath{\matV}{\bV}
\safemath{\matW}{\bW}
\safemath{\matX}{\bX}
\safemath{\matY}{\bY}
\safemath{\matZ}{\bZ}
\safemath{\matzero}{\bmzero}

\safemath{\matDelta}{\bDelta}
\safemath{\matLambda}{\bLambda}
\safemath{\matPhi}{\bPhi}
\safemath{\matSigma}{\bSigma}
\safemath{\matOmega}{\bOmega}
\safemath{\matTheta}{\bTheta}

\safemath{\matidentity}{\matI}
\safemath{\matone}{\matO}

\safemath{\rnda}{A}
\safemath{\rndb}{B}
\safemath{\rndc}{C}
\safemath{\rndd}{D}
\safemath{\rnde}{E}
\safemath{\rndf}{F}
\safemath{\rndg}{G}
\safemath{\rndh}{H}
\safemath{\rndi}{I}
\safemath{\rndj}{J}
\safemath{\rndk}{K}
\safemath{\rndl}{L}
\safemath{\rndm}{M}
\safemath{\rndn}{N}
\safemath{\rndo}{O}
\safemath{\rndp}{P}
\safemath{\rndq}{Q}
\safemath{\rndr}{R}
\safemath{\rnds}{S}
\safemath{\rndt}{T}
\safemath{\rndu}{U}
\safemath{\rndv}{V}
\safemath{\rndw}{W}
\safemath{\rndx}{X}
\safemath{\rndy}{Y}
\safemath{\rndz}{Z}

\safemath{\rveca}{\bimA}
\safemath{\rvecb}{\bimB}
\safemath{\rvecc}{\bimC}
\safemath{\rvecd}{\bimD}
\safemath{\rvece}{\bimE}
\safemath{\rvecf}{\bimF}
\safemath{\rvecg}{\bimG}
\safemath{\rvech}{\bimH}
\safemath{\rveci}{\bimI}
\safemath{\rvecj}{\bimJ}
\safemath{\rveck}{\bimK}
\safemath{\rvecl}{\bimL}
\safemath{\rvecm}{\bimM}
\safemath{\rvecn}{\bimN}
\safemath{\rveco}{\bomO}
\safemath{\rvecp}{\bimP}
\safemath{\rvecq}{\bimQ}
\safemath{\rvecr}{\bimR}
\safemath{\rvecs}{\bimS}
\safemath{\rvect}{\bimT}
\safemath{\rvecu}{\bimU}
\safemath{\rvecv}{\bimV}
\safemath{\rvecw}{\bimW}
\safemath{\rvecx}{\bimX}
\safemath{\rvecy}{\bimY}
\safemath{\rvecz}{\bimZ}

\safemath{\rvecxi}{\bmxi}
\safemath{\rveclambda}{\bmlambda}
\safemath{\rvecmu}{\bmmu}
\safemath{\rvectheta}{\bmtheta}
\safemath{\rvecphi}{\bmphi}

\safemath{\rmatA}{\bimA}
\safemath{\rmatB}{\bimB}
\safemath{\rmatC}{\bimC}
\safemath{\rmatD}{\bimD}
\safemath{\rmatE}{\bimE}
\safemath{\rmatF}{\bimF}
\safemath{\rmatG}{\bimG}
\safemath{\rmatH}{\bimH}
\safemath{\rmatI}{\bimI}
\safemath{\rmatJ}{\bimJ}
\safemath{\rmatK}{\bimK}
\safemath{\rmatL}{\bimL}
\safemath{\rmatM}{\bimM}
\safemath{\rmatN}{\bimN}
\safemath{\rmatO}{\bimO}
\safemath{\rmatP}{\bimP}
\safemath{\rmatQ}{\bimQ}
\safemath{\rmatR}{\bimR}
\safemath{\rmatS}{\bimS}
\safemath{\rmatT}{\bimT}
\safemath{\rmatU}{\bimU}
\safemath{\rmatV}{\bimV}
\safemath{\rmatW}{\bimW}
\safemath{\rmatX}{\bimX}
\safemath{\rmatY}{\bimY}
\safemath{\rmatZ}{\bimZ}

\safemath{\rmatDelta}{\bimDelta}
\safemath{\rmatLambda}{\bimLambda}
\safemath{\rmatPhi}{\bimPhi}
\safemath{\rmatSigma}{\bimSigma}
\safemath{\rmatOmega}{\bimOmega}
\safemath{\rmatTheta}{\bimTheta}

\usepackage{amssymb}
\usepackage{amsfonts}
\usepackage{mathrsfs}
\usepackage{xspace}
\usepackage{bm}
\usepackage{fancyref}
\usepackage{textcomp}

\usepackage{multirow}
\usepackage{stmaryrd}

\newenvironment{textbmatrix}{	\setlength{\arraycolsep}{2.5pt}%
								\big[\begin{matrix}}{\end{matrix}\big]%
								\raisebox{0.08ex}{\vphantom{M}}}

\def\be{\begin{equation}}
\def\ee{\end{equation}}
\def\een{\nonumber \end{equation}}
\def\mat{\begin{bmatrix}}
\def\emat{\end{bmatrix}}
\def\btm{\begin{textbmatrix}}
\def\etm{\end{textbmatrix}}

\def\ba#1\ea{\begin{align}#1\end{align}}
\def\bas#1\eas{\begin{align*}#1\end{align*}}
\def\bs#1\es{\begin{split}#1\end{split}} 
\def\bg#1\eg{\begin{gather}#1\end{gather}}
\def\bml#1\eml{\begin{multline}#1\end{multline}}
\def\bi#1\ei{\begin{itemize}#1\end{itemize}}

\safemath{\dirac}{\delta}					
\safemath{\krond}{\dirac}					

\safemath{\upto}{\uparrow}
\safemath{\downto}{\downarrow}
\safemath{\iu}{j}							
\safemath{\ev}{\lambda}						
\safemath{\hilseqspace}{l^{2}}				
\newcommand{\banachfunspace}[1]{\setL^{#1}}	
\safemath{\hilfunspace}{\banachfunspace{2}}	

\safemath{\SNR}{\textsf{SNR}} 				
\safemath{\PAR}{\textsf{PAR}} 				
\safemath{\No}{N_0}							
\safemath{\Es}{E_s}							
\safemath{\Eb}{E_b}							
\safemath{\EbNo}{\frac{\Eb}{\No}}
\safemath{\EsNo}{\frac{\Es}{\No}}

\DeclareMathOperator{\CHop}{\ensuremath{\opH}} 
\safemath{\tvir}{\rndh_{\CHop}}				
\safemath{\tvtf}{\rndl_{\CHop}}				
\safemath{\spf}{\rnds_{\CHop}}				
\safemath{\bff}{H_{\CHop}}					

\safemath{\ircf}{r_{h}}						
\safemath{\tftvcf}{r_{s}}					
\safemath{\tfcf}{r_{l}}						
\safemath{\bfcf}{r_{H}}						

\safemath{\tcorr}{c_h}						
\safemath{\scf}{c_{s}}						
\safemath{\tfcorr}{c_{l}}					
\safemath{\fcorr}{c_{H}}						

\safemath{\mi}{I}							
\safemath{\capacity}{C}						

\safemath{\normal}{\mathcal{N}}			
\safemath{\jpg}{\mathcal{CN}}			
\safemath{\mchain}{\leftrightarrow}		

\safemath{\dB}{\,\mathrm{dB}}
\safemath{\dBm}{\,\mathrm{dBm}}
\safemath{\Hz}{\,\mathrm{Hz}}
\safemath{\kHz}{\,\mathrm{kHz}}
\safemath{\MHz}{\,\mathrm{MHz}}
\safemath{\GHz}{\,\mathrm{GHz}}
\safemath{\s}{\,\mathrm{s}}
\safemath{\ms}{\,\mathrm{ms}}
\safemath{\mus}{\,\mathrm{\text{\textmu}s}}
\safemath{\ns}{\,\mathrm{ns}}
\safemath{\ps}{\,\mathrm{ps}}
\safemath{\meter}{\,\mathrm{m}}
\safemath{\mm}{\,\mathrm{mm}}
\safemath{\cm}{\,\mathrm{cm}}
\safemath{\m}{\,\mathrm{m}}
\safemath{\W}{\,\mathrm{W}}
\safemath{\mW}{\, \mathrm{mW}}
\safemath{\J}{\,\mathrm{J}}
\safemath{\K}{\,\mathrm{K}}
\safemath{\bit}{\,\mathrm{bit}}
\safemath{\nat}{\,\mathrm{nat}}

\safemath{\define}{\triangleq}			

\safemath{\equivalent}{\sim}
\safemath{\distas}{\sim}					
\safemath{\sdiff}{\Delta}				

\safemath{\reals}{\mathbb{R}}
\safemath{\positivereals}{\reals_{+}}
\safemath{\integers}{\mathbb{Z}}
\safemath{\posint}{\integers_{+}}
\safemath{\naturals}{\mathbb{N}}
\safemath{\posnaturals}{\naturals_{+}}
\safemath{\complexset}{\mathbb{C}}
\safemath{\rationals}{\mathbb{Q}}

\newcommand*{\fancyrefapplabelprefix}{app}		
\newcommand*{\fancyrefthmlabelprefix}{thm}		
\newcommand*{\fancyreflemlabelprefix}{lem}		
\newcommand*{\fancyrefcorlabelprefix}{cor}		
\newcommand*{\fancyrefdeflabelprefix}{def}		
\newcommand*{\fancyrefproplabelprefix}{prop}	
\newcommand*{\fancyrefobslabelprefix}{obs}		
\newcommand*{\fancyrefalglabelprefix}{alg}		
\newcommand*{\fancyrefasmlabelprefix}{asm}	    
\newcommand*{\fancyreftbllabelprefix}{tbl}	    

\frefformat{vario}{\fancyrefseclabelprefix}{Sec.~#1}
\frefformat{vario}{\fancyrefthmlabelprefix}{Thm.~#1}
\frefformat{vario}{\fancyreflemlabelprefix}{Lem.~#1}
\frefformat{vario}{\fancyrefcorlabelprefix}{Corr.~#1}
\frefformat{vario}{\fancyrefdeflabelprefix}{Def.~#1}
\frefformat{vario}{\fancyrefobslabelprefix}{Obs.~#1}
\frefformat{vario}{\fancyrefasmlabelprefix}{Ass.~#1}
\frefformat{vario}{\fancyreffiglabelprefix}{Fig.~#1}
\frefformat{vario}{\fancyrefapplabelprefix}{App.~#1} 
\frefformat{vario}{\fancyrefproplabelprefix}{Prop.~#1}
\frefformat{vario}{\fancyrefalglabelprefix}{Alg.~#1}
\frefformat{vario}{\fancyrefeqlabelprefix}{(#1)}
\frefformat{vario}{\fancyreftbllabelprefix}{Table~#1}

\safemath{\dictab}{[\,\dicta\,\,\dictb\,]}

\safemath{\ysig}{\bmy}
\safemath{\ysighat}{\hat{\ysig}}
\safemath{\ysigdim}{M}
\safemath{\xsig}{\bmx}
\safemath{\xsigdim}{N}
\safemath{\nx}{n_x}
\safemath{\zsig}{\bmz}
\safemath{\zsigdim}{\ysigdim}
\safemath{\rsig}{\bmr}
\safemath{\Adict}{\bA}
\safemath{\Adicttilde}{\widetilde{\Adict}}
\safemath{\Adictdim}{\outputdim\times\xsigdim}
\safemath{\avec}{\bma}
\safemath{\avectilde}{\tilde{\avec}}
\safemath{\Bdict}{\bB}
\safemath{\Bdicttilde}{\widetilde{\Bdict}}
\safemath{\Cdict}{\bC}
\safemath{\cvec}{\bmc}
\safemath{\Ddict}{\bD}
\safemath{\Ddictdim}{\ysigdim\times\xsigdim}
\safemath{\dvec}{\bmd}
\safemath{\Ddicttilde}{\widetilde{\bD}}
\safemath{\Bonb}{\bB}
\safemath{\bvec}{\bmb}
\safemath{\Bonbdim}{\ysigdim\times\ysigdim}
\safemath{\noise}{\bmn}
\safemath{\noisedim}{\ysigim}
\safemath{\err}{\bme}
\safemath{\errdim}{\ysigdim}
\safemath{\errset}{\setE}
\safemath{\nerr}{n_e}
\safemath{\delop}{\bP_\errset}
\safemath{\delopc}{\bP_{{\errset}^c}}

\safemath{\cplxi}{\imath}
\safemath{\cplxj}{\jmath}

\safemath{\dict}{\matD}
\safemath{\inputdim}{N}		
\safemath{\outputdim}{M}		
\safemath{\sparsity}{S}	
\safemath{\inputdimA}{{N_a}}	
\safemath{\inputdimB}{{N_b}}	
\safemath{\elemA}{{n_a}}	
\safemath{\elemB}{{n_b}}	
\safemath{\resA}{\matR_a}	
\safemath{\resB}{\matR_b}	
\safemath{\subD}{\matS} 
\safemath{\subA}{\matS_a} 
\safemath{\subB}{\matS_b} 
\safemath{\dicta}{\matA} 	
\safemath{\dictb}{\matB} 	
\safemath{\hollowS}{H}
\safemath{\hollowA}{H_a}
\safemath{\hollowB}{H_b}
\safemath{\cross}{Z}
\safemath{\coh}{\mu_d}			
\safemath{\coha}{\mu_a}			
\safemath{\cohb}{\mu_b}			
\safemath{\mubs}{\nu}	
\safemath{\cohm}{\mu_m} 
\safemath{\dictset}{\setD}	
\safemath{\dictsetp}{\dictset(\coh,\coha,\cohb)}	
\safemath{\dictsetgen}{\dictset_\text{gen}}
\safemath{\dictsetgenp}{\dictsetgen(\coh)}
\safemath{\dictsetonb}{\dictset_\text{onb}}
\safemath{\dictsetonbp}{\dictsetonb(\coh)}

\safemath{\leftside}{U}
\safemath{\rightsideA}{R_a}
\safemath{\rightsideB}{R_b}

\safemath{\indexS}{\setI_S} 

\safemath{\nb}{n_b}			
\safemath{\coeffa}{p_i}	
\safemath{\coeffb}{q_j}	
\safemath{\seta}{\setP}		
\safemath{\setb}{\setQ}     
\safemath{\setw}{\setW}	
\safemath{\setz}{\setZ}	
\safemath{\cola}{\veca}		
\safemath{\colb}{\vecb}		
\safemath{\cold}{\vecd}		
\safemath{\inputvec}{\vecx} 	
\safemath{\error}{\vece}	
\safemath{\noiseout}{\vecz} 	
\safemath{\inputvecel}{x}
\safemath{\inputveca}{\vecx_a}
\safemath{\inputvecb}{\vecx_b}
\safemath{\outputvec}{\vecy}	
\safemath{\lambdamin}{\lambda_{\mathrm{min}}}

\safemath{\elltwo}{\ell_2}
\safemath{\ellone}{\ell_1}
\safemath{\ellzero}{\ell_0}
\safemath{\ellinf}{\ell_\infty}
\safemath{\ellinftilde}{\ell_{\widetilde\infty}}
\safemath{\licard}{Z(\coh,\coha,\cohb)}
\safemath{\xsol}{\hat{x}}
\safemath{\xbord}{x_b}		
\safemath{\xstat}{x_s}		
\safemath{\xstatLone}{\tilde{x}_s}
\safemath{\order}{\mathcal{O}} 
\safemath{\scales}{\Theta} 
\safemath{\ones}{\mathbf{1}} 
\safemath{\zeroes}{\mathbf{0}} 
\safemath{\thlone}{\kappa(\coh,\cohb)} 
\safemath{\constoneA}{\delta} 
\safemath{\constoneB}{\epsilon} 
\safemath{\nlarge}{L}				   
\safemath{\sumlarge}{S_\nlarge}
\safemath{\maxlarger}{P_\nlarge}	   
\safemath{\Pzero}{\textrm{P0}}	
\safemath{\Pone}{\textrm{P1}}
\safemath{\vecfir}{\vecw}			 
\safemath{\vecsec}{\vecz}
\safemath{\elvecfir}{w}              
\safemath{\elvecsec}{z}				 
\safemath{\nlargefir}{n}
\safemath{\normout}{\gamma}
\safemath{\auxfun}{h}
\safemath{\supp}{\textrm{supp}}

\safemath{\indexa}{\ell}
\safemath{\indexb}{r}
\safemath{\indexc}{i}
\safemath{\indexd}{j}

\safemath{\project}{P}

\IEEEoverridecommandlockouts

\safemath{\Herm}{\textnormal{H}}
\begin{document}

\bstctlcite{IEEEexample:BSTcontrol}

\title{
A 105\,MS/s 4.18\,nJ/inference 0.765\,$\text{mm}^\text{2}$ Mixed-Signal Streaming ML Engine\\ for Wireless Baseband in 12LP+ FinFET}
\author{\IEEEauthorblockN{Oscar Casta\~neda, Lavinia Recchioni, Sueda Taner, Thomas Burger, J\'er\'emy Guichemerre, and Christoph Studer} \\[-0.3cm]
\thanks{This work has received funding from the Swiss State Secretariat for Education, Research, and Innovation (SERI) under the SwissChips initiative, and has also been supported in part through the Rigoletto project that has received funding from Chips Joint Undertaking (CHIPS-JU) under Grant Agreement 101194371. The authors would like to thank GlobalFoundries for providing silicon fabrication through the 12LP+ University Partnership Program. 
Contact author: O. Casta\~neda (e-mail: caoscar@ethz.ch)}
\IEEEauthorblockA{\emph{Department of Information Technology and Electrical Engineering, ETH Zurich, Switzerland}
} 
}

\maketitle

\begin{abstract}
We present SIXPPAC, the first streaming machine-learning engine for real-time integrated sensing and communication (ISAC) in wireless systems. SIXPPAC integrates two SAR ADCs and a digital binary-weight six-layer neural network~(NN). The 0.765\,mm$^\text{2}$ 12\,nm FinFET engine stores all NN parameters on chip, enabling sustained 105\,MS/s processing of 8b I/Q baseband samples at 0.8\,V. By deploying bipolar inner products using AND instead of XNOR gates, SIXPPAC consumes 4.18\,nJ/inference (0.42\,POPS/W) at 0.8\,V. We evaluate SIXPPAC on three example applications to demonstrate its versatility.
\end{abstract}


\section{Introduction}\label{sec:intro}

Machine learning (ML) is widely used in wireless communication systems, improving high-level physical-layer tasks such as channel estimation, data detection, forward error correction, and even positioning~\cite{zhang2020jetcas,attari2025}.
In contrast, ML remains largely unexplored for low-level baseband tasks that operate directly on received in-phase and quadrature (I/Q) samples.
Many such tasks (e.g., blind SNR estimation, channel busy rate detection, and time synchronization in frequency-selective channels) are in principle ill-posed, with no unique, optimal solution.
Practical approaches therefore exploit specific signal properties:
For instance, the M2M4 method~\cite{pauluzzi2000} for blind SNR estimation relies on the noise being Gaussian while the signal is not, whereas OFDM systems rely on a cyclic prefix to tolerate timing deviations within its guard interval~\cite{schmidl97a}.
These examples illustrate that classical methods rely on structural assumptions to yield practical solutions.
Thus, ML offers a promising alternative, as it can learn signal properties directly from data.

However, deploying ML in this setting is challenging:
Supporting low-level baseband tasks requires processing hundreds of millions of I/Q samples per second under strict latency constraints.
These tight requirements rule out conventional large neural networks~(NNs) due to their reliance on off-chip data movement.
Instead, these demands call for lightweight, domain-specific NNs that can be mapped to high-throughput spatial architectures with compact on-chip storage, while remaining expressive enough to support various baseband tasks.

\begin{figure}[t]
\centering
\includegraphics[width=0.935\linewidth]{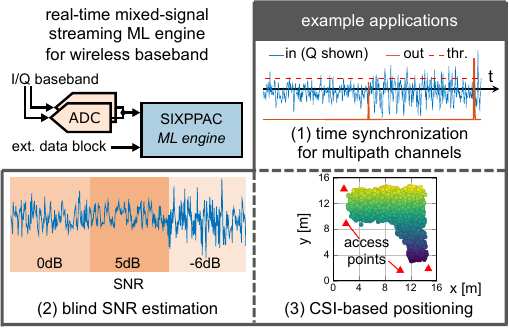}
  \caption{Overview of SIXPPAC ML engine with three example applications: I/Q samples from integrated ADCs are used for (1)~time synchronization and (2)~SNR estimation, while external CSI data is used for (3)~positioning.}
  \label{fig:intro}
\end{figure}

\subsubsection*{Contributions}
We present SIXPPAC (cf. \fref{fig:intro}), the first (to the best of our knowledge) streaming ML engine for real-time wireless baseband.
SIXPPAC implements a six-layer feedforward NN with binary weights, storing all weights, biases, and activations on chip to attain high throughput and low latency.
Two integrated 8b SAR ADCs digitize incoming I/Q baseband signals, eliminating digital input bottlenecks.
Besides continuous I/Q streaming operation, SIXPPAC supports block processing of multidimensional data.
We showcase three applications to demonstrate that SIXPPAC outperforms classical approaches (cf. \fref{sec:impl_per}) and enables ASIC efficiency for I/Q-sample-based integrated sensing and communication~(ISAC).


\section{VLSI Architecture} \label{sec:arch}

\begin{figure}[t]
\centering
\includegraphics[width=0.97\linewidth]{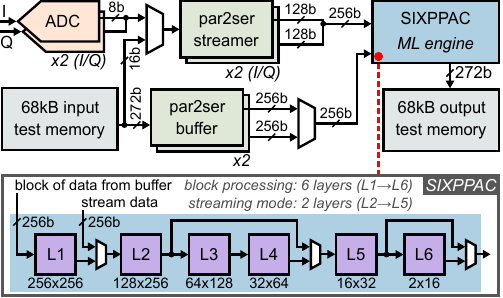}
  \caption{Top-level architecture of SIXPPAC. Each layer has dedicated hardware and all NN weights, biases, and activations are stored on chip.}
  \label{fig:arch}
\end{figure}

\fref{fig:arch} shows the top-level architecture of SIXPPAC.
The core engine is a six-layer perceptron with fully programmable weights and biases. 
Each fully-connected layer is implemented with dedicated hardware, enabling concurrent processing of consecutive input vectors.
SIXPPAC has two operation modes:
In the streaming mode, two layers process a window of 128 I/Q samples provided by a pair of integrated ADCs.
In the block-processing mode, all six layers process a 256-dimensional vector read from a test memory.
SIXPPAC operates bit-serially, processing one bit from all entries of the input vector per clock cycle.
Since its inputs are in bit-parallel format, parallel-to-serial converters are used.
SIXPPAC writes its results in bit-parallel format to an output test memory.

\subsection{Analog Frontend} \label{sec:arch_adc}

\begin{figure}[t]
\centering
\includegraphics[width=0.97\linewidth]{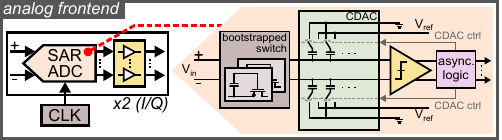}
  \caption{Details of the SAR ADC analog frontend integrated within SIXPPAC.}
  \label{fig:adc}
\end{figure}

\fref{fig:adc} details the two integrated 8b SAR ADCs, which reside in an independent power and clock domain, and buffer their outputs before delivery to the digital core.
The ADCs achieve 7.30\,ENOB at 105\,MS/s (cf. \fref{sec:impl}), enabling SIXPPAC to process 8b I/Q samples over a 105\,MHz RF bandwidth.
Thanks to SIXPPAC's bit-serial operation, higher sampling rates are possible when processing fewer input bits. For example, processing 1b per I/Q sample increases the sampling rate 8$\times$ to 840\,MS/s, where the ADCs provide a sufficient 5.71\,ENOB.

\subsection{ML Engine} \label{sec:arch_dig}

\begin{figure}[t]
\centering
\includegraphics[width=0.97\linewidth]{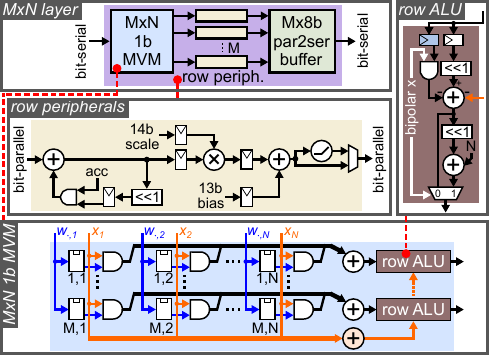}
  \caption{Hardware architecture for each SIXPPAC layer, consisting of a digital in-memory 1b MVM, row peripherals, and a parallel-to-serial buffer.\vspace{-0.3cm}}
  \label{fig:mvm}
\end{figure}

\fref{fig:mvm} shows the dedicated hardware for each NN layer.
Each layer has an $M\times N$ matrix-vector multiplier (MVM) built upon the digital in-memory compute (IMC) ``PPAC'' architecture from~\cite{castaneda2019ppac}.
The MVM multiplies a binary weight matrix~$\bW$ by a binary input vector~$\bmx$.
The 1b MVM outputs are further processed by row peripherals that (i)~accumulate the MVM results to support multi-bit inputs, (ii)~apply per-row 14b scale factors (to mitigate effects from coarse weight quantization) and 13b biases learned during off-chip training, and (iii)~perform linear or ReLU activation.
Since the MVM requires bit-serial inputs but produces bit-parallel outputs, the outputs of each layer are serialized before being fed to the next layer.

\subsection{AND-Based Bipolar Inner Product}

Prior NN accelerators (e.g., \cite{castaneda2019ppac,knag2021,lin2024,jia2022}) implement bipolar (HI/LO logic levels correspond to +1/-1, respectively)  multiplication using XNOR gates, as an XNOR gate outputs the same result when its inputs have the same logic level (or sign). 
Then, the inner product $\bmw^T\bmx$ between two bipolar binary vectors $\bmw$ and $\bmx$ of dimension $N$ can be computed as follows:

\begin{equation}
\label{eq:orig}
\bmw^T\bmx = 2 \sum_{i=1}^{N} w_i \odot x_i - N,
\end{equation}
where $\odot$ is the XNOR operator, and $w_i$ and $x_i$ take the value of their logic level (HI/LO). 
The sum operator $\sum$ treats HI as 1 and LO as 0. Therefore, the factor of 2 and bias of $-N$ correct for the mismatch arising from mapping a LO $x_i$ (or $w_i$) to 0 instead of its intended bipolar value of -1~\cite{castaneda2019ppac}.

With $\overline{x}$ being the logical negation of $x$, the XNOR operator~is
\begin{equation}
\label{eq:xnor}
w_i \odot x_i = w_i \cdot x_i + \overline{w_i}\cdot\overline{x_i},
\end{equation}
where the $\cdot$ and $+$ operators correspond to logical AND and OR, respectively.
However, since $w_i$ and $x_i$ take their logical values HI/LO, the AND~$\cdot$ operator is equivalent to multiplication with HI/LO$\equiv$1/0.
Likewise, since $w_i \cdot x_i$ and $\overline{w_i}\cdot\overline{x_i}$ cannot be simultaneously HI, the OR~$+$ operator can be treated as addition with HI/LO$\equiv$1/0.
Thus, we can re-express logical negation as $\overline{x} = 1-x$ and the XNOR operator~\fref{eq:xnor} as
\begin{equation}
\label{eq:xnor_new}
w_i \odot x_i = w_i \cdot x_i + (1-w_i)\cdot(1-x_i) = 2 w_i \cdot x_i - w_i - x_i + 1.
\end{equation}
Applying \fref{eq:xnor_new} to \fref{eq:orig}, we arrive at
\begin{equation}
\label{eq:and-based}
\bmw^T\bmx = 4 \sum_{i=1}^{N} w_i \cdot x_i - 2 \sum_{i=1}^{N} w_i - 2 \sum_{i=1}^{N} x_i + N,
\end{equation}
which, unlike \fref{eq:orig}, does not use XNOR $w_i \odot x_i$, but AND~$w_i \cdot x_i$.

\begin{figure}[t]
\centering
\subfigure[XNOR]{\includegraphics[width=.17\columnwidth]{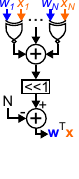}\label{fig:logic_xnor}}
\hfill
\subfigure[AND]{\includegraphics[width=.173\columnwidth]{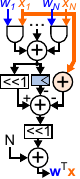}\label{fig:logic_and}}
\hfill
\subfigure[Energy and area comparison]{\includegraphics[width=.60\columnwidth]{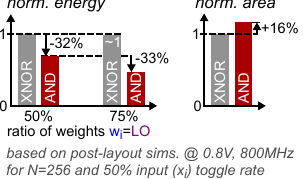}\label{fig:logic_comp}}
\caption{Block diagrams of 1b MVMs that perform bipolar ($\pm$1) inner products using (a) XNOR vs. (b) AND-based logic. (c) Compared to previous works~\cite{castaneda2019ppac,knag2021,lin2024,jia2022} using XNOR gates, our novel AND-based approach saves energy at a modest area increase, while supporting unipolar (0,1) inputs as well.}
  \label{fig:and}
\end{figure}

\fref{fig:and} shows both XNOR- and AND-based inner-product hardware implementations, and compares their energy and area.
Our AND-based inner product reduces energy by 32\% while increasing area by only 16\% compared to the traditional XNOR approach:
Lower energy is achieved as a LO weight prevents any toggling activity at its AND's output.
The area increase is due to the additional hardware (colored in Figs. \ref{fig:and} and \ref{fig:mvm}) required to compute $\sum_i x_i$ and $\sum_i w_i$ in \fref{eq:and-based}.
This additional hardware has a negligible impact on the overall energy, as the input vector~$\bmx$ is the same for a given layer, and as weights~$\bmw$ remain static over several inferences.
Hence, the dynamic energy overheads of computing $\sum_i x_i$ and $\sum_i w_i$ are diluted across the IMC array and multiple inferences, respectively.

In addition, the energy of AND-based MVMs decreases as more weights are LO: When 75\% of the weights are LO, energy reduces by 33\% compared to when 50\% of the weights are LO, as more AND outputs do not toggle.
The worst-case energy occurs at 50\% LO weights, since, if a row has more HI than LO weights, all weights can be flipped and the row's scale factor sign inverted without affecting the result.
In stark contrast, the energy of the conventional XNOR approach remains practically constant regardless of the number of LO weights.
Finally, our AND approach naturally extends to inner products with unipolar~(HI/LO $\equiv$ 1/0) inputs, enabling SIXPPAC to support both multi-bit bipolar and two's complement inputs.


\section{Implementation Results}\label{sec:impl}

\begin{figure}[t]
\begin{minipage}{0.47\linewidth}
\centering
\includegraphics[width=0.99\linewidth]{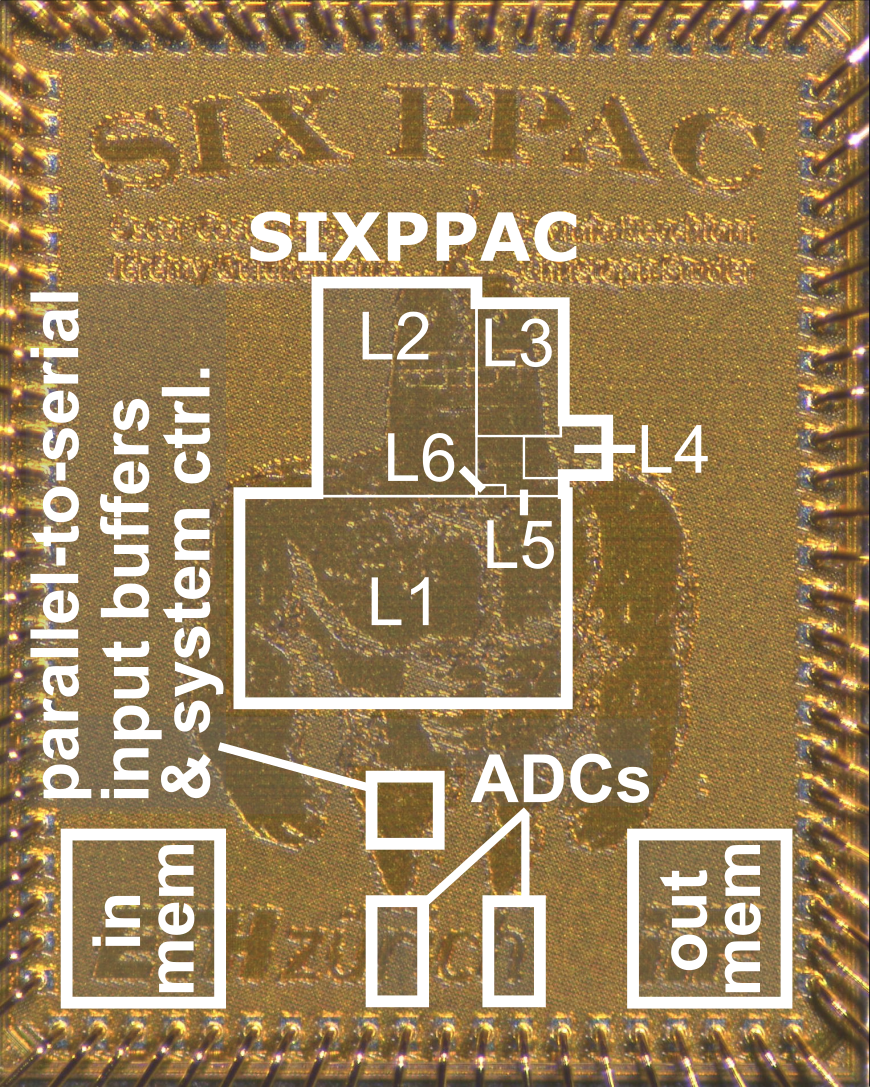}
  \caption{Micrograph of 2\,mm$\times$2.5\,mm 12LP+ FinFET SIXPPAC die.}
  \label{fig:chip}
  \end{minipage}
  \hfill
\begin{minipage}{0.48\linewidth}
\centering
\subfigure[Measurement setup]{\includegraphics[width=.9\columnwidth]{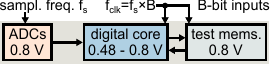}\label{fig:vfs_setup}}
\subfigure[Measurement results]{\includegraphics[width=.965\columnwidth]{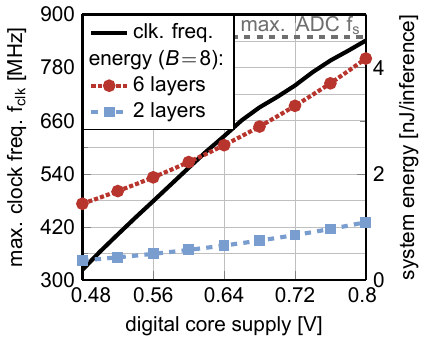}\label{fig:vfs_results}}
  \caption{Measurement (a)~setup and (b)~results at 300\,K room temperature.}
  \label{fig:vfs}
  \end{minipage}
\end{figure}

\begin{table}[t]
  \setlength{\tabcolsep}{4pt}
  \caption{Measurements of Implemented SAR ADCs}
  \label{tab:adc}
  \vspace{-0.2cm}
  \centering
  \scalebox{0.99}{
    \begin{tabular}{@{}lc|lccc@{}}
      \toprule
      Technology [nm] & 12 & Sampling rate $f_\text{s}$ [MS/s] & 105 & 600 & 850 \\
      Resolution [bit] & 8 & ENOB @ $f_\text{in}\cong f_\text{s}/2$ [bit] & 7.30 & 7.26 & 5.71 \\
      Supply [V] & 0.8 & SNDR @ $f_\text{in}\cong f_\text{s}/2$ [dB] & 45.7 & 45.5 & 36.1 \\
      Active area [mm$^\text{2}$] & 0.144 &  Total power [mW] & 0.618 & 2.29 & 2.63 \\
      \bottomrule
    \end{tabular}
  }
\end{table}

\fref{fig:chip} shows the SIXPPAC die fabricated in 12LP+ FinFET, integrating the ML engine and two SAR ADCs in 0.765\,mm$^\text{2}$.
Table~\ref{tab:adc} and \fref{fig:vfs} detail our measurements for the ADCs and digital core, respectively.
At 0.8\,V, the digital core operates at 840\,MHz, processing 105\,MS/s for 8b inputs, a rate supported by the ADCs.
The digital core achieves its minimum energy at 0.48\,V, while the ADCs are always supplied with 0.8\,V.

\begin{figure}[t]
\centering
\subfigure[Time sync.: Total error rate]{\includegraphics[width=.47\linewidth]{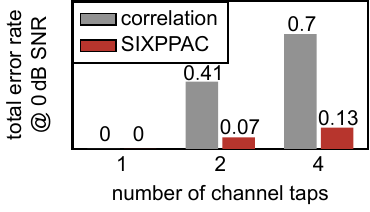}\label{fig:appres_syncter}}
\hfill
\subfigure[Time sync.: 4-tap ch. ROC]{\includegraphics[width=.475\linewidth]{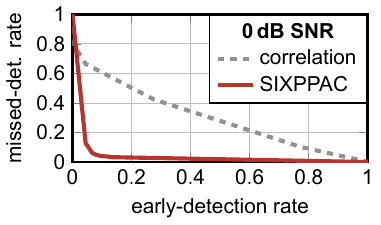}\label{fig:appres_syncroc}}
\subfigure[Blind SNR estimation]{\includegraphics[width=.475\linewidth]{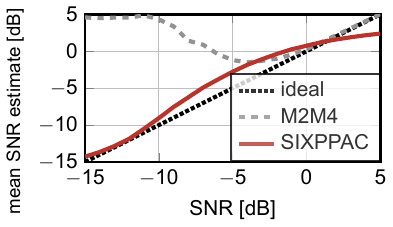}\label{fig:appres_snr}}
\hfill
\subfigure[CSI-based positioning]{\includegraphics[width=.465\linewidth]{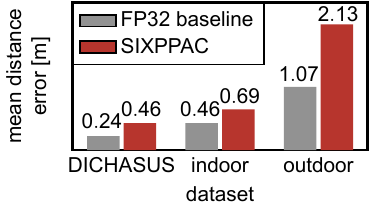}\label{fig:appres_pos}}
\caption{SIXPPAC performance for three applications: (a-b) time synchronization at 0\,dB SNR, (c) blind SNR estimation, and (d) CSI-based positioning (the DICHASUS dataset~\cite{euchner2021} covers a 12$\times$12\,m$^\text{2}$ area, the indoor dataset a 9.2$\times$26\,m$^\text{2}$ area, and the outdoor dataset an 81$\times$119\,m$^\text{2}$ area).}
  \label{fig:appres}
\end{figure}

\subsection{Performance for Example Applications} \label{sec:impl_per}

\fref{fig:appres} demonstrates SIXPPAC's versatility through three applications.
For time synchronization, SIXPPAC processes blocks of 128 I/Q samples from the ADCs to detect WiFi preambles without channel knowledge.
This task is traditionally tackled by computing the correlation between the received samples and the known preamble signal to detect when this correlation exceeds a certain threshold.
As mentioned in \fref{sec:intro}, small deviations in time synchronization can be tolerated in OFDM systems~\cite{schmidl97a}.
However, to highlight the capabilities of SIXPPAC, our results in Figs.~\ref{fig:appres_syncter} and \ref{fig:appres_syncroc} treat any deviation as a synchronization error.
Under this strict metric, correlation remains error-free for frequency-flat channels (i.e., channels with 1 tap), but is significantly outperformed by our ML-based SIXPPAC for frequency-selective channels.

For blind SNR estimation, SIXPPAC estimates SNR directly from blocks of 128 I/Q samples captured by the ADCs.
Traditionally, this task is performed using the moment-based M2M4 method~\cite{pauluzzi2000}, which fails under some conditions.
\fref{fig:appres_snr} shows that low SNR is one of such conditions, where SIXPPAC delivers robust performance.

For positioning, SIXPPAC applies block processing to downlink channel-state information (CSI) from distributed multi-antenna access points.
This task is typically tackled using ML and floating-point~(FP) arithmetic~\cite{attari2025}.
\fref{fig:appres_pos} shows that, although SIXPPAC attains a higher positioning error than a 32b FP baseline, it still achieves sub-2.5\,m accuracy with binary weights, which enable a much more efficient hardware implementation.
These three examples demonstrate that SIXPPAC's binary-weight NN provides sufficient expressivity, while its programmable weights enable additional ISAC applications.

\begin{figure}[t]
\centering
\includegraphics[width=0.948\linewidth]{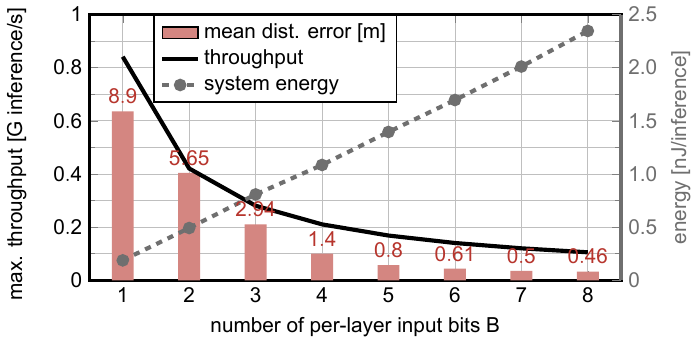}
  \caption{Impact of per-layer input resolution $B$ on throughput, energy, and positioning performance on the DICHASUS dataset~\cite{euchner2021} at 0.8\,V supply.}
  \label{fig:bitserial}
\end{figure}

\begin{figure}[t]
\centering
\includegraphics[width=0.948\linewidth]{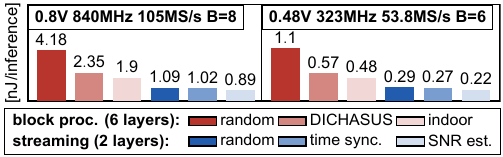}
  \caption{Energy (including ADC energy) per inference for different applications; ``random'' corresponds to 50\% LO weights and 50\% input toggle rate.}
  \label{fig:energy}
\end{figure}

\fref{fig:bitserial} shows how precision can be traded for higher processing rate and lower energy thanks to SIXPPAC's bit-serial design:
With the DICHASUS CSI positioning dataset~\cite{euchner2021}, using 6b instead of 8b inputs increases throughput by 33\% and reduces energy by 28\%, at the cost of a 15\,cm increase in average positioning error.
\fref{fig:energy} shows energy (including ADCs) for different tasks:
Random weights and inputs consume the most energy, while real-world applications benefit from natural weight and input sparsity.
At 0.8\,V and 105\,MS/s with 8b inputs, mean energy is 1.9\,nJ/inference (max. 4.18\,nJ/inference); at 0.48\,V and 53.8\,MS/s with 6b inputs, mean energy drops 3.9$\times$.

\subsection{Comparison with State-of-the-Art} \label{sec:impl_comp}

\begin{table}
\setlength{\tabcolsep}{4pt}
    \caption{Comparison with State-of-the-Art IMC Inference Accelerators That Support Bipolar Inner Products}
    \label{tab:comp}
    \vspace{-0.3cm}
    \begin{center}
    \scalebox{0.99}{
    \begin{tabular}{@{}lcc|cccc@{}}
        \toprule
        ~ & \multicolumn{2}{c}{This work}  & Knag & Lin & Jia & Desoli \\
        ~ & \multicolumn{2}{c}{SIXPPAC}  & \cite{knag2021} & \cite{lin2024} & \cite{jia2022} & \cite{desoli2023} \\
        \midrule
        Input ADCs & \multicolumn{2}{c}{\textbf{yes}} & no & no & no$^\textit{a}$ & no \\
        \midrule
        Computation & \multicolumn{2}{c}{digital,} & digital, & digital, & analog+ & digital, \\
        paradigm & \multicolumn{2}{c}{\textbf{exact}} & \textbf{exact} & approx. & digital & \textbf{exact} \\
        \midrule
        Bipolar inner product & \multicolumn{2}{c}{\textbf{AND-based}} & XNOR & XNOR & XNOR & no$^\textit{b}$ \\
        \midrule
        IMC array size [kb] & \multicolumn{2}{c}{107} & 128 & 16 & 4\,608 & 2\,048 \\
        Technology [nm] &  \multicolumn{2}{c}{12} & 10 & 28 & 16 & 18 \\
        Area [mm$^\text{2}$] & \multicolumn{2}{c}{0.765} & 0.39 & 0.033 & 25 & 4.2 \\
        Weight prec. [bit] & 1 & 1 & 1 & 1 & 4 & 4\\
        In/act. prec. [bit] & 8 & 6 & 1 & 1 & 4 & 4\\
        Core supply [V] & 0.8 & 0.48 & 0.75 & 0.9 & 0.8 & 1.0\\
        Clock freq. [MHz] & 840 & 323 & 622 & 280 & 200 & -\\
        Power [mW] & 439 & 59 & 607 & - & - & -\\
        \midrule
        ADC rate [MS/s] & \textbf{105} & 53.8 & - & - & 20$^\textit{a}$ & -\\
        Latency [ns] & \textbf{87} & 189 & - & - & - & -\\
        TOPS$^\textit{c}$ & 183 & 70.5 & 163 & 9.18 & 189 & \textbf{229}$^\textit{d}$ \\
        TOPS/mm$^\text{2}$ & 240 & 92.1 & \textbf{418} & 278 & 7.55 & 54.5 \\
        POPS/W$^\textit{c,e}$ & 0.42 & 1.2 & 0.27 & 0.57$^\textit{f}$ & \textbf{1.9} & 0.2$^\textit{g}$ \\
        \bottomrule
    \end{tabular}
    } \vspace{-0.1cm}
    \end{center}
    {\scriptsize$^\textit{a}$Includes ADCs for internal analog compute; these ADCs cannot be used for external analog inputs. $^\textit{b}$Only supports unipolar/two's complement inner products. $^\textit{c}$Numbers normalized to 1b$\times$1b compute. $^\textit{d}$Throughput for 1b$\times$1b compute extracted from \cite[Fig. 16.7.7]{desoli2023}. $^\textit{e}$Energy efficiency at 50\% input toggle rate, except for \cite{knag2021,jia2022}, which use CIFAR10 and a CNN toggle activity, respectively. $^\textit{f}$Energy efficiency at 50\% input toggle rate extracted from \cite[Fig. 15.c]{lin2024}. $^\textit{g}$Energy efficiency at 1.0\,V supply not provided in~\cite{desoli2023}; instead, we show energy efficiency at 0.525\,V, extracted from \cite[Fig. 16.7.5]{desoli2023}.}
\end{table}

SIXPPAC is, to the best of our knowledge, the first design integrating the input ADCs required to enable real-time baseband ML processing.
The closest design to SIXPPAC is \cite{attari2025}, an unfabricated ASIP with a CNN accelerator for communications and positioning that delivers 390 positioning/s.
In contrast, SIXPPAC achieves 105\,M\,positioning/s with 87\,ns latency in a comparable area.
Table~\ref{tab:comp} compares SIXPPAC to state-of-the-art IMC inference accelerators that support bipolar inner products.
SIXPPAC delivers the highest throughput~(TOPS) among sub-1\,mm$^\text{2}$ designs, with competitive compute density even when including the ADC area.
Its AND-based compute enables best-in-class energy efficiency (POPS/W) for exact bipolar inner products.
Alternatives \cite{lin2024,jia2022} improve energy efficiency through approximate or analog compute, which preclude bit-accurate results.
In particular, analog compute enables \cite{jia2022} to achieve higher energy efficiency at low compute density; the all-digital SIXPPAC offers a more balanced point in this trade-off.
Complete NN inference systems like \cite{desoli2023} suffer from low compute density and energy efficiency due to data-movement overheads.
In contrast, thanks to AND-based IMC and a fixed NN architecture with fully programmable parameters, SIXPPAC approaches the performance of inference accelerators \cite{knag2021,lin2024,jia2022} while implementing complete I/Q-sample-based ISAC tasks with tight real-time constraints.


\section{Conclusions}

We have presented SIXPPAC, the first mixed-signal ML engine for wireless baseband that directly operates on a stream of I/Q samples.
By integrating SAR ADCs with an AND-based NN engine with fully on-chip storage, SIXPPAC unlocks ASIC hardware efficiency for various I/Q-sample-based ISAC tasks.
SIXPPAC outperforms traditional approaches in time synchronization and blind SNR estimation, and enables high-rate positioning, demonstrating that real-time ML processing of I/Q baseband samples is not only feasible, but advantageous.



\end{document}